\documentclass[preprint]{aastex62}
\usepackage{graphicx}
\hypersetup{linkcolor=red,citecolor=green,filecolor=cyan,urlcolor=magenta}

\received{July 30, 2020}
\revised{Aug 11, 2019}
\accepted{Aug 18, 2020}
\submitjournal{ApJL}

\shorttitle{Propagating Density Fluctuations in Solar Coronal Holes}
\shortauthors{Cho et al.}

\begin{document}

\title{Accelerating and Supersonic Density Fluctuations in Coronal Hole Plumes: Signature of Nascent Solar Winds}

\correspondingauthor{Yong-Jae Moon}
\email{moonyj@khu.ac.kr}

\author[0000-0001-7514-8171]{Il-Hyun Cho}
\affiliation{Department of Astronomy and Space Science, Kyung Hee University, Yongin, 17104, Korea}

\author[0000-0001-6423-8286]{Valery M. Nakariakov}
\affiliation{School of Space Research, Kyung Hee University, Yongin, 17104, Korea}
\affiliation{Centre for Fusion, Space and Astrophysics, Department of Physics, University of Warwick, CV4 7AL, UK}
\affiliation{Special Astrophysical Observatory, Russian Academy of Sciences, St. Petersburg, 196140, Russia}

\author[0000-0001-6216-6944]{Yong-Jae Moon}
\affiliation{Department of Astronomy and Space Science, Kyung Hee University, Yongin, 17104, Korea}
\affiliation{School of Space Research, Kyung Hee University, Yongin, 17104, Korea}

\author[0000-0001-6412-5556]{Jin-Yi Lee}
\affiliation{Department of Astronomy and Space Science, Kyung Hee University, Yongin, 17104, Korea}

\author[0000-0003-1459-3057]{Dae Jung Yu}
\affiliation{School of Space Research, Kyung Hee University, Yongin, 17104, Korea}

\author[0000-0003-2161-9606]{Kyung-Suk Cho}
\affiliation{Space Science Division, Korea Astronomy and Space Science Institute, Daejeon 34055, Korea}
\affiliation{Department of Astronomy and Space Science, University of Science and Technology, Daejeon 34055, Korea}

\author[0000-0001-9982-2175]{Vasyl Yurchyshyn}
\affiliation{Big Bear Solar Observatory, New Jersey Institute of Technology, 40386 North Shore Lane, Big Bear City, CA 92314-9672, USA}

\author[0000-0002-9300-8073]{Harim Lee}
\affiliation{Department of Astronomy and Space Science, Kyung Hee University, Yongin, 17104, Korea}

\begin{abstract}
Slow magnetoacoustic waves in a static background provide a seismological tool to probe the solar atmosphere in the analytic frame.
By analyzing the spatiotemporal variation of the electron number density of plume structure in coronal holes above the limb
for a given temperature,
we find that the density perturbations accelerate with supersonic speeds in the distance range from 1.02 to 1.23 solar radii.
We interpret them as slow magnetoacoustic waves propagating at about the sound speed with accelerating subsonic flows.
The average sonic height of the subsonic flows is calculated to be 1.27 solar radii.
The mass flux of the subsonic flows is estimated to be 44.1$\%$ relative to the global solar wind.
Hence, the subsonic flow is likely to be the nascent solar wind.
In other words, the evolution of the nascent solar wind in plumes at the low corona is quantified for the first time from imaging observations.
Based on the interpretation, propagating density perturbations present in plumes could be used as a seismological probe
of the gradually accelerating solar wind.
\end{abstract}

\section{Introduction} \label{intro}
Slow magnetoacoustic waves are useful seismological tool to probe the solar atmosphere \citep[e.g.,][]{2017ApJ...837L..11C, 2019ApJ...871L..14C}.
MHD waves propagating in a flowing background with a constant speed are faster than phase speeds of the waves in a static medium due to the Doppler effect,
which was predicted by theories \citep{1992SoPh..138..233G, 1996A&A...311..311N}
and observations \citep{2011ApJ...728..147C, 2011SoPh..272..119F, 2020ApJ...893...78D}.
A wave propagation in a flowing background with a non-constant speed may not be analyzed analytically, but can be explored in a simulation \citep{2020ApJ...893...64G}.

Solar plumes are thin and ray-like structures rooted above networks and extended up to at least 30 solar radii
\citep{1997SoPh..175..393D, 2001ApJ...546..569D}.
Plumes are known to be cooler and denser than their surrounding interplumes \citep{2015LRSP...12....7P}.
It was found that a plume in extreme ultraviolet (EUV) bands disappears due to a density reduction rather than temperature decrease \citep{2014ApJ...793...86P}.
It was also found that EUV intensities are enhanced above the enhanced spicular activity \citep{2019Sci...366..890S}.
These structures are thought to be magnetic tubes that guide MHD waves
\citep{2006RSPTA.364..473N, 2011SSRv..158..267B, 2015LRSP...12....7P}
which were observed as periodically propagating intensity disturbances in various wavelength bands
\citep{1997ApJ...491L.111O, 1998ApJ...501L.217D, 1999ApJ...514..441O, 2009A&A...503L..25W, 2010ApJ...718...11G,
2011A&A...528L...4K, 2012A&A...546A..93G, 2014ApJ...789..118K, 2018ApJ...868..149K},
and/or mass-flows \citep{2010A&A...510L...2M, 2011ApJ...736..130T, 2014ApJ...793...86P}.

In this study, we provide evidence of wave propagations in an accelerating background with a subsonic speed in plume structures.
For this, a propagation speed of density fluctuations in plume structures is estimated and compared with the sound speed obtained from the
electron temperature given by the differential emission measure (DEM).
In Section 2, we describe data and method to distinguish the plume structures in a plume line of sight (LOS).
In Section 3, we perform the least-square fitting to the evolution of density fluctuation by the second-order polynomial
and explore the property of background flow.
Finally, we summarize and discuss our results.

\section{Data and Method}\label{dm}
\subsection{Plume structures in the plume LOS}
We use narrow-band filtergram images at 94, 131, 171, 193, 211, 335 {\AA}
taken by the Atmospheric Imaging Assembly (AIA) \citep{2012SoPh..275...17L}
on board the Solar Dynamics Observatory (SDO) \citep{2012SoPh..275....3P} on 2017-Jan-03 00:00 UT -- 24:00 UT.
During the observation day, the entire Sun was very quiet.
Each image is rotated based on the solar P angle
and re-sized to have the pixel resolution of 0.600 arcseconds at the distance of 1.496$\times$10$^{8}$ km
which is the reference distance given in data.
The intensity is scaled according to the changes in pixel resolution and the disk size.

\begin{figure*}\centering
\includegraphics[scale=0.68, angle=90]{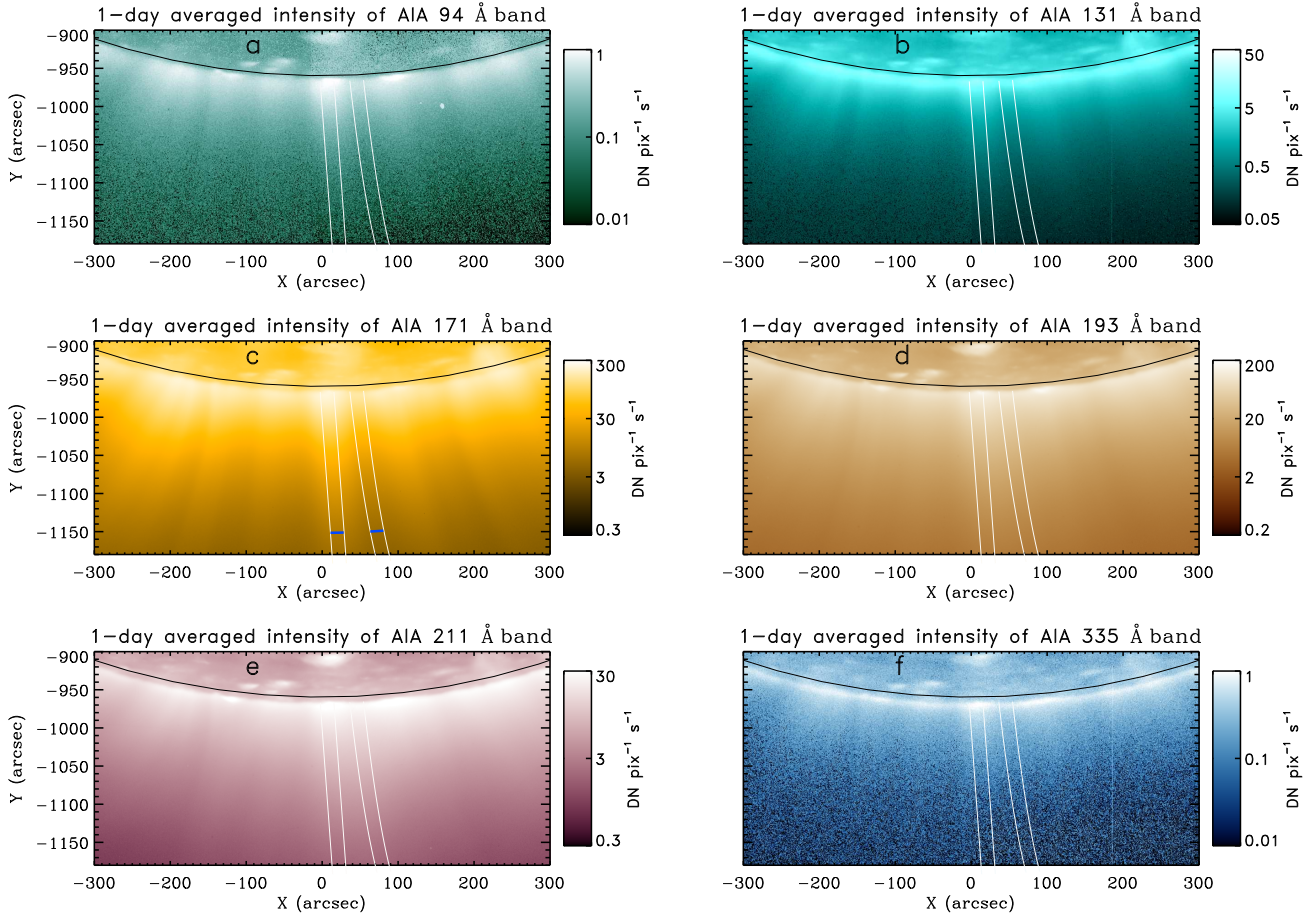}
\caption{\small
One-day averaged intensities of 94, 131, 171, 193, 211, 335 {\AA} bands taken by the SDO/AIA for the south polar region (a -- f).
Two white lines on the left and right in each panel are boundaries of plume LOS and interplume LOS.
From these region, intensities for plume and interplume LOSs are constructed as a function of heliocentric distance and time.
The blue horizontal lines indicate the height of 1.2 solar radii.
}
\label{met.fig1}\end{figure*}

Plumes are embedded in interplume background.
To minimize the effects from interplume emission on the estimation of the density and temperature,
we define the plume line of sight (LOS) and interplume LOS separately, as in Figure~\ref{met.fig1}.
We define slits which indicate the boundaries of plume and interplume LOSs
based on the 1-day averaged intensity of the AIA 171 {\AA} band (Figure~\ref{met.fig1}c).
The slits on the intensity images for the other five bands are the same with that of the intensity image of 171 {\AA}.
The plume was inclined to $\sim$4$^\circ$ relative to the direction normal to the solar surface.
The interplume was also inclined, but the axis looks to be curved.
Both the positions of straight and curved lines are determined by 1st- and 2nd- order polynomial fittings
from several locations that were visually determined.
Along the slits,
the intensities of the (inter)plume LOS between boundaries at the (right)left are averaged for a given height and frame.
For example, the average intensity of 171 {\AA} band on the plume LOS at the distance of 1.2 solar radii is determined
from intensities along the positions indicated by the blue line on the left in Figure~\ref{met.fig1}c,
and that on the interplume is from the positions indicated by the blue line on the right.
Note that slit distances in the plume and interplume LOSs are different from each other at a given height.
The distance in our study represents the heliocentric distance corresponding to the inclined slit distance of the plume LOS.

\subsection{Differential emission measure}
The DEM represents the amount of emission of plasma,
and gives an electron number density for a given temperature and length of the LOS.
The intensity of the filtergram with narrow ranges of wavelength on EUV
taken by the SDO/AIA ($I_i$) can be modeled as
$\int_T R_i(T) \mathrm{DEM}(T) dT$,
where $I_i$ and $R_i$ represent intensity and temperature response for a certain channel.
Temperature response functions are slightly different for different abundances \citep[e.g.,][]{2017ApJ...844....3L}.
It is likely that abundance enhancements are not able to be built up in an open magnetic field structure of coronal holes.
Hence, we use the photospheric abundance \citep{2011SoPh..268..255C}
which gives the temperature response of 171 {\AA} at around 0.8 MK to be lower $\sim$2 times
compared to that from the conventional coronal one \citep{1992PhyS...46..202F}.

\begin{figure*}\centering
\includegraphics[scale=0.7]{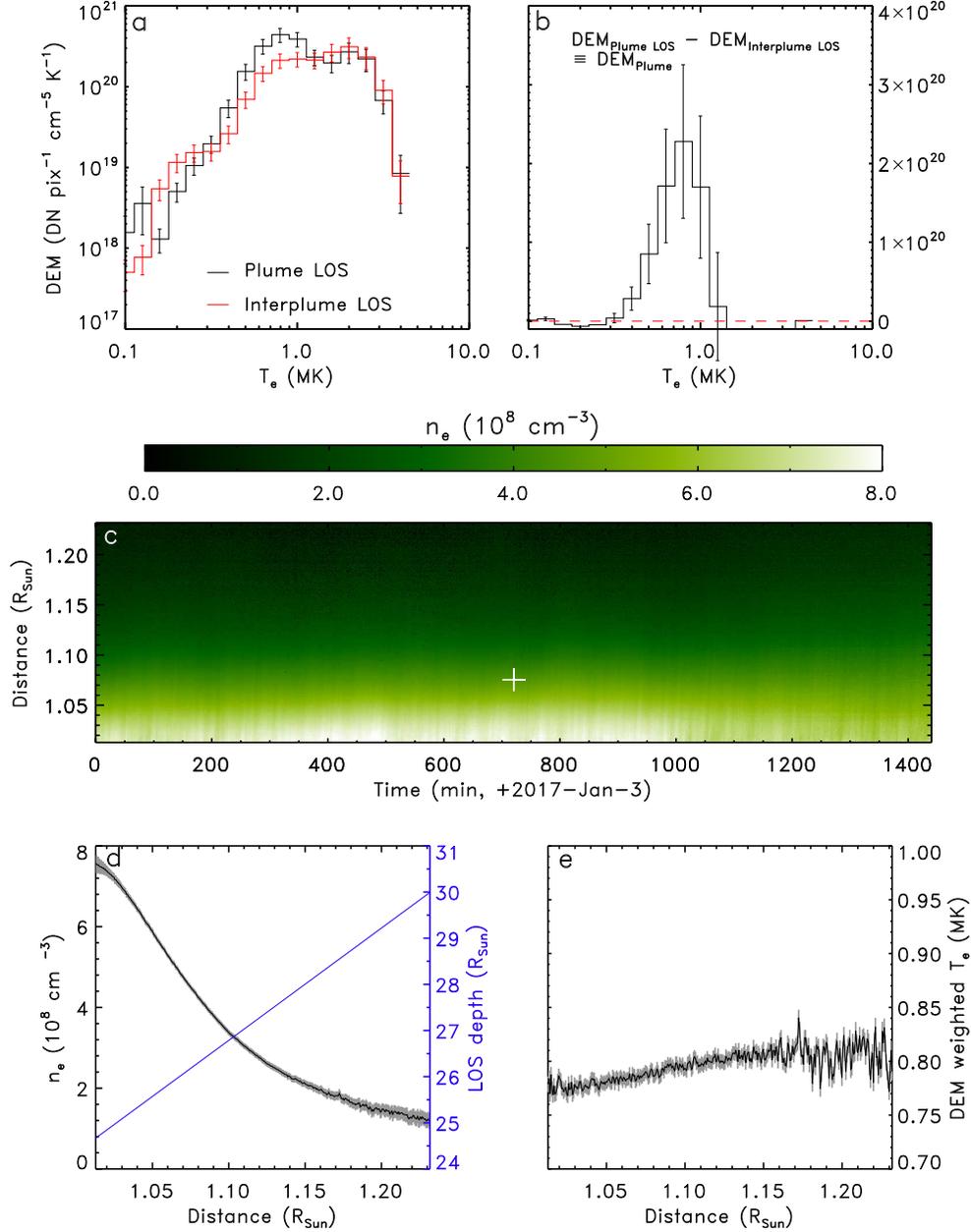}
\caption{
Examples of DEMs for plume LOS (black) and interplume LOS (red) at an arbitrary time and distance (a), their difference (b),
the map of electron number density (c), temporal averages of the electron number density and temperature as a function of distance (d, e).
The vertical bars in panel a represent $\pm 1\sigma_\mathrm{DEM_\mathrm{Plume \ LOS}}$ and $\pm 1 \sigma_\mathrm{DEM_\mathrm{Interplume \ LOS}}$.
The vertical bars in panel b represent $\pm$1$\sigma_{\Delta \mathrm{DEM}}$
defined by $\sqrt{ \sigma_{\mathrm{DEM}^2_\mathrm{Plume \ LOS}} + \sigma_{\mathrm{DEM}^2_\mathrm{Interplume \ LOS}} }$.
The red-dashed line in panel b indicates the zero DEM.
The gray area in panel d and e represent $\pm 1\sigma_{n_e}$ and $\pm 1 \sigma_{T_e}$. 
The blue-solid line in panel d is the diameter of a single plume.
}
\label{met.fig3}\end{figure*}

To deriv the DEM, we apply the recently developed method, the Solar Iterative Temperature Emission Solver (SITES) \citep{2019SoPh..294..135M}
for a given pixel on the time-distance images constructed on the plume LOS and the interplume LOS.
The method calculate a DEM directly from the observed intensities and fractional temperature response functions.
As a result, we obtain DEM$_\mathrm{Plume \ LOS}(t, h, T)$ and DEM$_\mathrm{Interplume \ LOS}(t, h, T)$,
where $t$, $h$, and $T$ are the time, height, and temperature bin, respectively.
DEMs are 1-min averaged to enhance the signal-to-noise ratios.
From this, we define the DEM of plume structures
$\mathrm{DEM}_\mathrm{Plume}$ as $\mathrm{DEM}_\mathrm{Plume \ LOS} - \mathrm{DEM}_\mathrm{Interplume \ LOS}$.
A snapshot of DEM$_\mathrm{Plume \ LOS}$, DEM$_\mathrm{Interplume \ LOS}$, and DEM$_\mathrm{Plume}$
are presented in Figures~\ref{met.fig3}a and~\ref{met.fig3}b, which are taken from the position indicated by the cross in Figure~\ref{met.fig3}c.
As shown in Figure~\ref{met.fig3}a, both DEMs have two bumps at around 0.8 MK and 2 MK, but the latter bumps are identical in both LOSs.
It was found that the temperature of the equatorial coronal holes is $\sim$0.9 MK,
but becomes higher if the region of interest includes outer quiet regions \citep{2020SoPh..295....6S}.
The off limb measurement certainly includes emissions from the quiet region at different heights.
Hence, we believe that the temperature of former bumps is likely to be the typical value of the plume in coronal holes.
This result is well explained with an assumption when both the plume and interplume LOSs includes plume structures and $\sim$2 MK backgrounds,
but the interplume LOS includes less number of plume structures.
Hence, the subtraction of DEM$_\mathrm{Interplume \ LOS}$ from DEM$_\mathrm{Plume}$
can minimize a contribution to plume emissions from the background,
but also reduces emissions from plumes along the plume LOS.

The emission measure, $\mathrm{EM}_\mathrm{Plume}(t, h)$, is defined as $\int_T \mathrm{DEM}_\mathrm{Plume}(t, h, T) dT$.
The electron number density, $n_e(t, h)$, is defined as $\sqrt{ \frac{\mathrm{EM}_\mathrm{Plume}(t, h)}{dl(h)} }$,
and presented in Figure~\ref{met.fig3}c.
The LOS length of plumes, $dl(h)$, is set to be the length of chord ($2h\tan{\theta}$) for a single plume,
where $h$ is the height, $\theta$ is half of the angular width of a plume (1$^{\circ}$), which corresponds to 24 -- 30 Mm.
The calculated number density seems to be consistent with the measurement from on-disk coronal holes \citep{2020SoPh..295....6S}.
The electron temperature, $T_e(t, h)$, is defined as $\frac{\int_T \mathrm{DEM}_\mathrm{Plume} T dT}{\int_T \mathrm{DEM}_\mathrm{Plume} dT }$.
The temporal averages of $n_e$ and $T_e$ for a given distance are presented in Figures~\ref{met.fig3}d and~\ref{met.fig3}e.
These quantities are used for the estimation of the mass flux and sound speed.

\section{Results}
\subsection{Evolution of propagating density disturbances}
\begin{figure*}\centering
\includegraphics[scale=0.65]{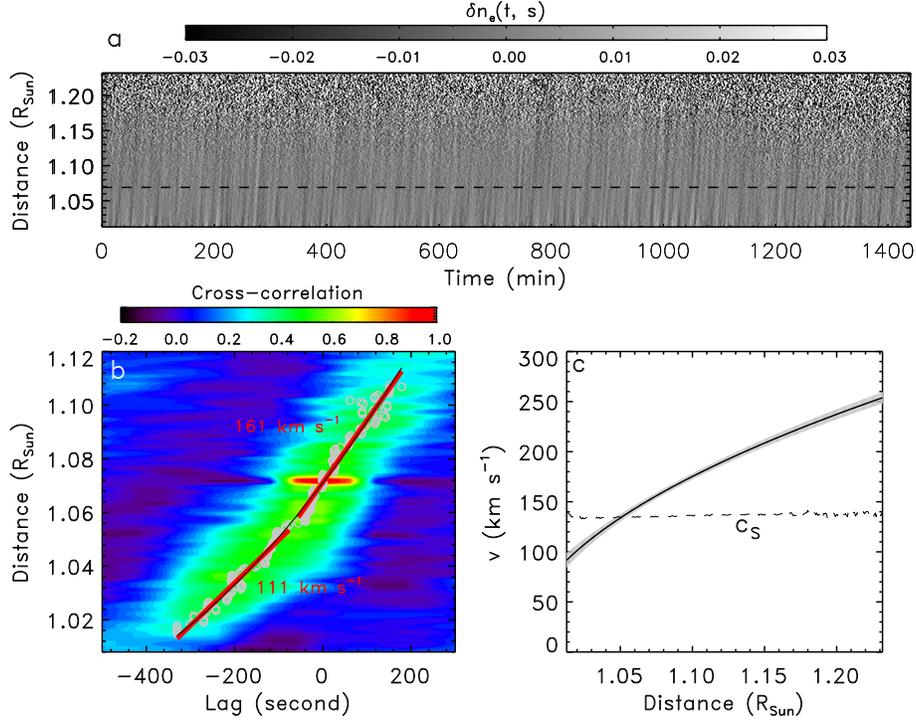}
\caption{\small
Perturbations of electron number density ($\delta n_e \equiv d(n_e/<n_e>_t-1$)) as a function of time and heliocentric distance (a),
the average cross-correlations obtained from 95 sub-images, as a function of lag and distance (b),
and the average speed of the density perturbations as a function of distance (c).
In panel b, the gray circle indicates the weighted mean of the correlation for a given distance.
Two red-solid lines represent the linear least-square fits for the distances as a function of lags.
The thin black-solid line is the 2nd order polynomial fit.
In panel c, the speed and its range of error are represented by the black and gray colors.
The dashed line represents the sound speed calculated from the temporally averaged temperature for a given distance (see Figure~\ref{met.fig3}e).
}
\label{res.fig1}\end{figure*}

We analyze $\delta n_e (t, h)$ defined by the electron number density ($n_e$) sequentially subtracted by the previous one, for a given height (Figure~\ref{res.fig1}a).
We perform the median smoothing with 3 minutes by 3 pixels ($\sim$1.3 Mm) to suppress short-term fluctuations.
By visual inspection, there are many propagating quasi-periodic density perturbations during one-day.
We divide the time-distance image into 95 sub-images every 15 minutes having a temporal range $\pm$15 minutes (31 minutes).
For each sub-image, we calculate the lagged cross-correlations between the profile at the distance of 50 Mm and profiles for different distances.
The average cross-correlation is presented in Figure~\ref{res.fig1}b.
The positive and negative lags represent that the profiles at different distances lead and trail the profile at the distance of 50 Mm, respectively,
hence the migration of the lag showing maximum correlations from negative to positive indicates the upward propagation.

In Figure~\ref{res.fig1}b, we plot the weighted-mean lag for a given distance (gray circle).
It is clearly shown that the instantaneous slope of distance evolution ($h(t)$) are different from different times (see red solid lines).
We perform the linear least-square fittings for the distances at low and high altitudes,
and found that the speeds are 111 km s$^{-1}$ and 161 km s$^{-1}$, respectively.
Hence, h(t) is likely to accelerate.
To quantify the evolution, $h(t)$ were fitted with the 2nd-order polynomial as a function of lag time.
As a result, the evolution of the propagation of perturbations is described by a constant acceleration model.
The acceleration ($a$) is calculated to be 183 $\pm$ 12 m s$^{-2}$.
The initial speed ($v_0$) at zero height ($h_0$) is found to be 67 km s$^{-1}$.

In Figure~\ref{res.fig1}c, we plot the fitted speed ($v$) as a function of distance together with the sound speed ($c_\mathrm{S}$).
The speed is given by $\sqrt{v^2_0 + 2a(h-h_0)}$
and its error $\sqrt{v_0^2 \delta v_0^2 + (h-h_0)^2 \delta a^2 + \frac{a^2(\delta h^2 + \delta h_0^2)}{v_0^2 + 2a(h-h_0)}}$,
where $\delta v_0$, $\delta a$, $\delta h_0$ are the errors of the fitting parameters,
and $\delta h$ is taken to be the standard deviation of residuals between $h$ and the observed distance.
The speed is compared with the sound speed which could be the propagation speed of slow magnetoacoustic waves in a static medium of low plasma-$\beta$.
The sound speed ($c_\mathrm{S}$) is $\sqrt{\frac{\gamma k_\mathrm{B}T}{\mu m_{H}}}$,
and equivalent to 90.9 $\sqrt{\frac{\gamma <T_e(t,h)>_{1 \mathrm{MK}}}{\mu}}$ km s$^{-1}$,
where $<T_e(t,h)>_{1 \mathrm{MK}}$ is the temporal average of temperature divided by 1 MK for a given distance as shown in Figure~\ref{met.fig3}e,
$\gamma (=1.67)$ is the adiabatic index, $k_\mathrm{B}$ is the Boltzmann constant, $m_\mathrm{H}$ is the proton mass, and $\mu (=0.6)$ is the mean molecular weight.
It is shown that the speed of the density perturbation becomes faster than the sound speed from $\sim$1.05 solar radii ($\sim$35 Mm)
and has an excess of $\sim$115 km s$^{-1}$ relative to the sound speed at 1.23 solar radii ($\sim$160 Mm)
(see a difference between black line and dashed line in Figure~\ref{res.fig1}c).
The excess speed seems to be consistent with radial speeds derived by the Doppler dimming technique \citep{2003ApJ...589..623G, 2003ApJ...588..566T}.
Hence, the excess speed in our study is likely to be the speed of flowing background.

\begin{figure*}\centering
\includegraphics[scale=0.5]{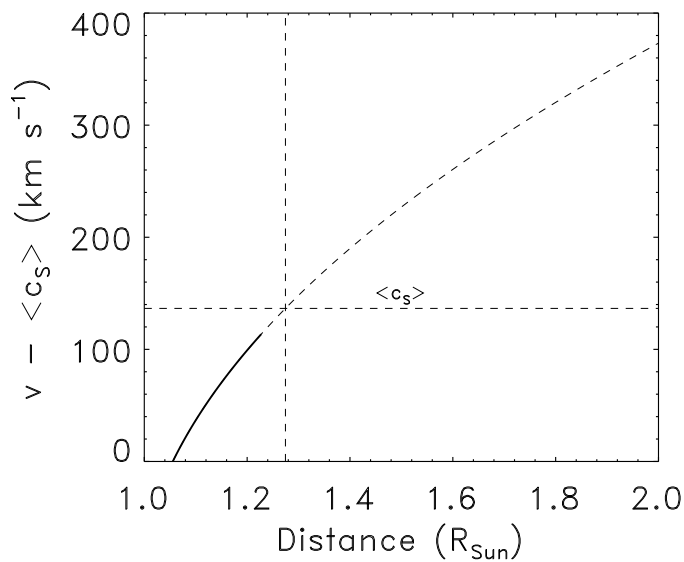}
\caption{\small
The average speed is extrapolated up to 2 solar radii using the fitting parameters and then
subtracted by the mean sound speed which corresponds to the phase speed of slow waves.
The horozontal-dashed line indicates the sound speed.
The vertical-solid line indicates the distance where the extrapolated speed becomes supersonic.
}
\label{res.fig2}\end{figure*}

In Figure~\ref{res.fig2}, we present the flow speed defined as the observed speed after subtracting off the mean sound speed, which is assumed to be wave speed.
Interestingly, the distance where the flow speed becomes supersonic is 1.27 solar radii when extrapolated using the fitting parameters.
This distance is lower than sonic heights of solar winds \citep{2019ApJ...881L..36T, 2020ApJ...893...64G}.
This may because the extrapolation is based on the constant acceleration motion,
which may not adequately describe complex dynamic evolution of solar wind such as deceleration at low altitude \citep{2017ApJ...846...86B}.

\subsection{Mass flux}
\begin{figure*}\centering
\includegraphics[scale=0.7]{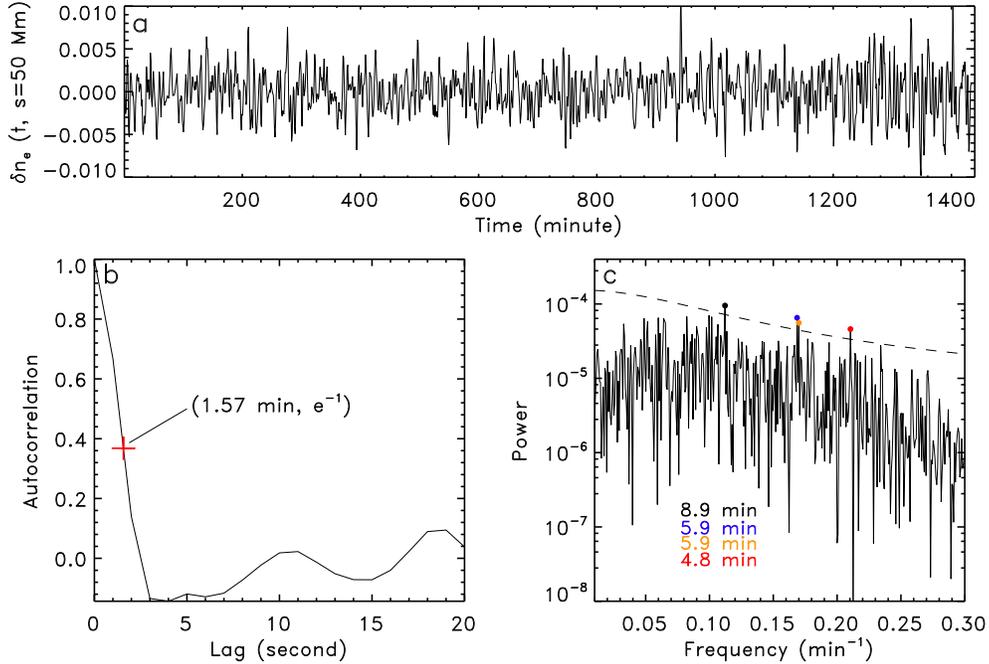}
\caption{
Perturbations of the electron number density at the distance of 50 Mm as indicated in~\ref{res.fig1}a (a),
and the corresponding autocorrelation (b) and the Fourier power (c).
The red cross indicates the decorrelation time (e-folding time).
The solid line in panel c represents 99.9\% significance level of the red noise.
Four colored circles represent the powers higher than the significance level.
Their peak peirods range from $\sim$5 -- $\sim$9 min.
}
\label{res.fig3}\end{figure*}

We apply spectral analysis to the density profile at the distance of 50 Mm as indicated in Figure~\ref{res.fig1}a.
We assume that the profile is embedded in red noise
because a perturbed medium at a certain time might be influenced by previous perturbations via dissipation or heating.
This may result in a frequency dependent power which is to be an additional noise.
A red noise is defined as $\frac{\sigma^2 (1-\rho^2)}{(1 - 2\rho \cos{\frac{f}{f_N}} + \rho^2)}$ \citep{2002CG.....28..421S}
where $\sigma$ is the standard deviation of the density profile in Figure~\ref{res.fig3}a,
$\rho$ is the autoregressive parameter of the autoregressive process of the order 1,
$f$ is the frequency (min$^{-1}$),
and $f_N$ is the Nyquist frequency, respectively.
The autoregressive parameter is defined by $e^{-\frac{\Delta t}{\tau_d}}$,
where $\Delta t$ is the sampling interval and 
$\tau_d$ is the decorrelation time which makes the autocorrelation to be $\frac{1}{e}$ as shown in Figure~\ref{res.fig3}b.
This noise follows the chi-square distribution with two degrees of freedom.
It is shown that the Fourier power at 4.8, 5.9, 8.9 minutes are above the 99\% noise level.
The observed periods will be used to estimate a temporal filling factor $(f_\mathrm{T}$).

The mass flux is defined as $4\pi d^2 \mu m_p n_e v f_\mathrm{T} f_\mathrm{CH} f_\mathrm{S}$ \citep{2014Sci...346A.315T},
where $d$ is the heliocentric distance,
$\mu$ is the mean molecular weight,
$m_p$ is the proton mass,
$n_e$ is the electron number density,
$v$ is the flow speed,
$f_\mathrm{T}$ is the temporal filling factor,
$f_\mathrm{CH}$ is the fractional area of the coronal hole,
$f_\mathrm{S}$ is the spatial filling factor.
In Figure~\ref{res.fig1}b, the perturbation is observed from -300 seconds to 300 seconds, hence the lifetime is at least 10 minutes.
The temporal filling factor, defined by the ratio of the lifetime ($\sim$10 minutes) to the period (5 -- 9 minutes), could be taken as the unity.
We use $f_\mathrm{CH}=0.05$ and $f_\mathrm{S}=0.1$,
indicating that plumes occupy 10\% of a coronal hole and the coronal hole covers 5\% area of the solar surface.
The mass flux is calculated to be
5.6$\times$11$^{11}$ g s$^{-1}$ (8.8$\times$10$^{-15}$ M$_{\odot}$ yr$^{-1}$),
if we apply $n_e = 2.1 \times 10^8$ cm$^{-3}$ and
$v = 66.5$ km s$^{-1}$ at the height of 100 Mm ($d \sim$ 1.144 solar radii)
(see Figure~\ref{met.fig3}d and Figure~\ref{res.fig2}).
This value corresponds to $\sim$44.1\% of the global solar wind \citep{2011MNRAS.417.2592C}.

\section{Summary and Discussion} \label{sm}
In this study,
we analyzed the kinematics of perturbations of the electron number density in plume structures above the limb,
as a function of time and heliocentric distance,
and find that the density perturbations are accelerating up to supersonic speeds for a given temperature.
We interpreted them as slow magnetoacoustic waves in a low plasma-$\beta$ background
which is flowing with subsonic speeds and exhibiting acceleration.
The acceleration of the subsonic flows is estimated to be 183 $\pm$ 12 m$^{-2}$ in the distance range from 1.02 to 1.23 solar radii.
The extrapolated sonic height is calculated to be 1.27 solar radii,
lower than sonic heights of solar winds ($\sim$2 solar radii) \citep{2019ApJ...881L..36T, 2020ApJ...893...64G}.
The discrepancy may be explained if solar winds decelerate within $\sim$1.5 solar radii and gently reaccelerate \citep{2017ApJ...846...86B}.
The mass flux corresponds 44.1\% to the global solar wind.
Hence, the flowing background is likely to be nascent solar winds.

To our knowledge, this is the first direct measurement of the solar wind speed in plumes from 1.02 to 1.23 solar radii from imaging observations.
Our measurement may help to constrain solar wind models at the low corona.
A slow wave in an isothermal plume could be used as a seismological probe of the gradually accelerating solar wind.
Our observation can support the simulation showing that wave signatures in the presence of solar wind are responsible for
propagating intensity features observed in the high corona up to $\sim$30 solar radii \citep{2020ApJ...893...64G},
which were ubiquitously observed in the coronagraphic images \citep{2018PhRvL.121g5101C, 2018ApJ...862...18D}.

If the density perturbations are repeated supersonic solar winds, the mass flux corresponds to 134.6\% on the global solar wind.
The repetition periods are in the narrow range from $\sim$5 to $\sim$9 minutes (Figure~\ref{res.fig3}c).
Hence, periodic sources are required.
If periodic magnetic reconnections are the sources \citep{2015ApJ...806..172S}, the flow speeds are Alfv\'{e}nic.
However, the observed speed seems to be sub-Alfv\'{e}nic.
Note that the typical Alfv\'{e}n speed in the low corona is over 600 km s$^{-1}$ \citep{2013A&A...556A.124T}.

The apparent variation of the phase speed could also be connected with the variation of
the polytropic index $\gamma$, and hence the effective sound speed with height in an isothermal and static
plasma, caused by the misbalance of heating and cooling processes \citep{2019PhPl...26h2113Z}.
A robust measurment of $\gamma$ as a function of height would be helpful to examine the possibility,
but such measurement seems only to be allowed on-disk where the signal-to-noise is high \citep[e.g.,][]{2018ApJ...868..149K}.
Coronal holes are possibly in nonequilibrium ionization (NEI) states \citep[e.g.,][]{2013SSRv..178..271B}.
It is shown that the measured plasma density and temperature could be affected by NEI in a rapidly heated system \citep[e.g.,][]{2019ApJ...879..111L},
while the NEI significantly affects the FIP and abundance in a coronal hole \citep{2019ApJ...882..154S}.
Possible effects of NEI modulated by a MHD wave were explored through a foward modeling \citep{2019ApJ...882..154S}.
An attempt to fomulate MHD waves under a NEI condition have been performed only recently \citep{2019FrASS...6...39B},
which potentially could provide a tool for interpreting observations.

\acknowledgments
We appreciate helpful comments from an anonymous reviewer, which improve the original manuscript.
The SDO data is (partly) provided by the Korean Data Center (KDC) for SDO
in the Korea Astronomy and Space Science Institute (KASI)
in cooperation with NASA/SDO and the AIA, EVE, and HMI science teams.
This work is supported by KASI under the R\&Dprogram
'Development of a Solar Coronagraph on International Space Station' (Project No. 2020-1-850-07)
supervised by the Ministry of Science, ICT and Future Planning,
the BK 21 plus program funded by the Korean Government,
the Basic Science Research Program through the National Research Foundation (NRF) of Korea
(grant No. NRF-2020R1I1A0107814) funded by the Ministry of Education,
and also the Research Program (grant No. NRF-2019R1C1C1006033, NRF-2019R1C1C1004778, NRF-2019R1A2C1002634)
funded by the Ministry of Science, ICT and Future Planning.
This work is also supported by
the Institute for Information \& communications Technology Promotion (IITP) grant funded by the Korean government (2018-0-0142).
V.M.N.acknowledges support from the STFC consolidated grant ST/T000252/1.
V.Y. acknowledges support from NSF AST-1614457, AFOSR FA9550-19-1-0040, and NASA80NSSC17K0016, 80NSSC19K0257, and 80NSSC20K0025 grants

\end{document}